\documentstyle[11pt,epsfig]{article}
\date{}
\begin{document}
\title{{\bf Evolutionary quantum cosmology in a gauge-fixed picture }}
\author{Babak Vakili\thanks{%
email: b-vakili@iauc.ac.ir}\\\\
{\small {\it Department of Physics, Azad University of
Chalous,}}\\{\small {\it P.O. Box 46615-397, Chalous, Iran}}}
\maketitle

\begin{abstract}
We study the classical and quantum models of a flat
Friedmann-Robertson-Walker (FRW) space-time, coupled to a perfect
fluid, in the context of the consensus and a gauge-fixed
Lagrangian frameworks. It is shown that, either in the usual or in
the gauge-fixed actions, the evolution of the Universe based on
the classical cosmology represents a late time power law
expansion, coming from a big-bang singularity in which the scale
factor goes to zero for the standard matter, and tending towards a
big-rip singularity in which the scale factor diverges for the
phantom fluid. We then employ the familiar canonical quantization
procedure in the given cosmological setting to find the
cosmological wave functions in the corresponding minisuperspace.
Using a gauge-fixed (reduced) Lagrangian, we show that, it may
lead to a Schr\"{o}dinger equation for the quantum-mechanical
description of the model under consideration, the eigenfunctions
of which can be used to construct the time dependent wave function
of the Universe. We use the resulting wave function in order to
investigate the possibility of the avoidance of classical
singularities due to quantum effects by means of the many-worlds
and ontological interpretation of quantum cosmology.\vspace{5mm}\noindent

\end{abstract}
\section{Introduction}
The standard model of relativistic cosmology has its origin in the
general theory of relativity. As is well known, standard
cosmological models based on classical general relativity have no
convincingly precise answer to the question of the initial
conditions from which the Universe has evolved. This can be traced
to the fact that these models suffer from the presence of an
initial singularity, the so-called big-bang singularity. Indeed,
there are various forms of singularity theorems in general
relativity \cite{Haw} which show that quite reasonable assumptions
lead to at least one consequence which is physically unacceptable.
Any hope of dealing with such singularities would be in the
development of a concomitant and conducive quantum theory of
gravity \cite{Dew}. In the absence of a full theory of quantum
gravity, it would be useful to describe the quantum states of the
Universe within the context of quantum cosmology. In this
formalism which is based on the canonical quantization procedure,
one first freezes a large number of degrees of freedom and then
quantizes the remaining ones. The quantum state of the Universe is
then described by a wave function in the minisuperspace, a
function of the $3$-geometry and matter fields presented in the
theory, satisfying the Wheeler-DeWitt equation \cite{Kief}. An
important feature of this wave function is that unlike the case of
ordinary quantum mechanics, the wave function in quantum cosmology
is time independent. This is indeed a reflect of the fact that
general relativity is a parameterized theory in the sense that its
(Einstein-Hilbert) action is invariant under time
reparameterization. Because the existence of the "problem of
time", that is, time evolution is lost in the dynamics of the wave
function, the canonical formulation of general relativity leads to
a constrained system and its Hamiltonian is a superposition of
some constraints, the so-called Hamiltonian and momentum
constraints. A possible way to overcome this problem is that one
first solves the equation of constraint to obtain a set of genuine
canonical variables with which one can construct a reduced
Hamiltonian. In this kind of time reparameterization, the
equations of motion are obtained from the reduced physical
Hamiltonian and describe the evolution of the system with respect
to the selected time parameter \cite{Am}.

An important ingredient in any model theory related to the
quantization of a cosmological model is the choice of the
quantization procedure used to quantized the system. As mentioned
above, the most widely used method has traditionally been the
canonical quantization method based on the  Wheeler-DeWitt
equation which is nothing but the application of the Hamiltonian
constraint to the wave function of the Universe. However, one may
solve the constraint before using it in the theory and, in
particular, before quantizing the system. Then, one can quantize
the reduced system in the same manner as in elementary quantum
mechanics. If we do so, since there are no constraints in the
reduced phase space, we are led to a Schr\"{o}dinger type equation
where a time reparameterization in terms of various dynamical
variables can be done before quantization \cite{Is}. Although such
kinds of gauge fixing lead to different time parameters, the
physical behavior of the system under consideration remains
invariant under the corresponding gauge transformations. There is
also another type of gauge fixing proposed in \cite{Mal}, and this
is removing the gauge freedom from the theory at the level of
Lagrangian. In this method, instead of the usual Einstein-Hilbert
Lagrangian one can construct a Lagrangian theory for the
gravitational system free of the traditional gauge freedoms and
show that the correct dynamics of the system is recovered  while
the equations of motion are different.

In this paper, we first deal with a spatially flat FRW cosmology
with a perfect fluid as its matter field. Classically, we show
that the corresponding dynamical system resulting from the
Einstein-Hilbert action has a compatible set of solutions and
constraints. Quantum mechanically, we employ the familiar
canonical quantization procedure in this cosmological setting and
investigate the behavior of the resulting wave function either in
the semiclassical and quantum regimes. We then consider the
problem at hand in the context of a gauge-fixed Lagrangian
proposed in \cite {Mal}. We see that the examined action is free
of the usual Hamiltonian constraint and thus it may lead to the
identification of a time parameter for the corresponding dynamical
system. Moreover, this formalism gives rise to a Schr\"{o}dinger
equation for the quantum-mechanical description of the model under
consideration. Finally, we use the resulting wave function in
order to investigate the possibility of the avoidance of classical
singularities due to quantum effects by means of the many-worlds
and ontological interpretation of quantum cosmology.
\section{Perfect fluid cosmology}
In this section, we start by briefly studying the ordinary,
spatially flat FRW model, where the metric is given by
\begin{eqnarray}\label{A}
ds^2=-N^2(t)dt^2+a^2(t)(dx^2+dy^2+dz^2),
\end{eqnarray}with $N(t)$ and $a(t)$ being the lapse function and the scale
factor, respectively. Also, we assume that a perfect fluid with
which the action of the theory is augmented plays the role of the
matter part of the model. The Einstein-Hilbert Lagrangian with a
general energy density $V(a)$ becomes \cite{Khos}
\begin{eqnarray}\label{B}
{\cal{L}}&=&\sqrt{-g}(R[g]-V(a))\nonumber\\
&=&-6N^{-1}a\dot{a}^2- Na^3 V(a),
\end{eqnarray}where $R[g]$ is the Ricci scalar and in the second line the total derivative term has been
ignored. For a perfect fluid where its pressure $p$ is linked to
its energy density $\rho$ by a barotropic Equation of State (EoS)
\begin{equation}\label{C}
p=\omega \rho,\end{equation}the function $V(a)$ becomes
$V(a)=M_{\omega}a^{-3(\omega+1)}$. Here, $M_{\omega}$ is a model
dependent constant and $-1\leq \omega \leq 1$ is the EoS
parameter. Therefore, by considering the above action as
representing a dynamical system in which the scale factor $a$
plays the role of the dynamical variable, we obtain the following
pointlike Lagrangian
\begin{equation}\label{D}
{\cal{L}}=-6N^{-1}a\dot{a}^2-
NM_{\omega}a^{-3\omega}.\end{equation}

\subsection{The classical model}
The classical dynamics are governed by the Euler-Lagrange equation
$\frac{d}{dt}\frac{\partial {\cal L}}{\partial
\dot{a}}-\frac{\partial {\cal L}}{\partial a}=0$, that is,
\begin{equation}\label{E}
4a\ddot{a}+2\dot{a}^2+N\omega
M_{\omega}a^{-3\omega-1}=0.\end{equation}As is well known the
cosmological model, in view of the concerning issue of time, has
been rather general and of course under-determined. Before trying
to solve this equation we must decide on a choice of time in the
theory. The under determinacy problem at the classical level may
be removed by using the gauge freedom via fixing the gauge. For
example, we can work in the gauge $N=1$, which usually is chosen
in classical cosmological models and is called the cosmic time
gauge. To proceed, we consider the gauge $N=1$ and assume that the
scale factor depends on the cosmic time as
\begin{equation}\label{F}
a(t)=(At-\tau_0)^{\beta},\end{equation}where $A$, $\tau_0$, and
$\beta$ are some constants. Using this ansatz in Eq. (\ref{E}), it
is straightforward to obtain
\begin{equation}\label{G}
2\beta-2=-(3\omega+1)\beta,\hspace{0.5cm}2\beta
(2-3\beta)A^2-\omega M_{\omega}=0,\end{equation} where their
solutions (for $\omega \neq -1$) can be written as
\begin{equation}\label{I}
\beta=\frac{2}{3(\omega+1)},\hspace{0.5cm}A=\sqrt{\frac{3M_{\omega}}{8}}(\omega+1).\end{equation}
Therefore, the classical cosmology exhibits a power law expansion
with the scale factor
\begin{equation}\label{J}
a(t)=\left[\sqrt{\frac{3M_{\omega}}{8}}(\omega+1)
t-\tau_0\right]^{\frac{2}{3(\omega+1)}}.\end{equation} For $\omega
=-1$, the perfect fluid plays the role of a cosmological constant
and in this case we can directly integrate the Eq. (\ref{E}) to
obtain
\begin{equation}\label{J1}
a(t)=a_0\exp \left(\sqrt{\frac{M_{-1}}{6}}t\right).\end{equation}
As a double check, one may obtain the above solutions from the
Einstein equations. Indeed, assuming the full Einstein equations
to hold, this implies that the Hamiltonian corresponding to the
Lagrangian (\ref{D}) must vanish, that is
\begin{equation}\label{K}
-6a\dot{a}^2+M_{\omega}a^{-3\omega}=0.\end{equation}It is easy to
check that the solutions (\ref{J}) and (\ref{J1}) automatically
satisfy the Hamiltonian constraint (\ref{K}), and thus we have a
compatible system of solutions and constraint. The evolution of
the Universe based on (\ref{J}) begins with a big-bang singularity
at $t=\sqrt{\frac{8}{3M_{\omega}}}\frac{\tau_0}{\omega+1}$ and,
for $\omega >-1$, follows the power law expansion $a(t)\sim
t^{\frac{2}{3(\omega+1)}}$ at the late time of cosmic evolution.
We can also compute the deceleration parameter
$q=-\frac{a\ddot{a}}{\dot{a}^2}$ for this model. As is well known
the deceleration parameter indicates by how much the expansion of
the Universe is slowing down. If the expansion is speeding up, for
which there appears to be some recent evidence, then this
parameter will be negative. A simple calculation shows that at
late time limit we have $q \sim \frac{3\omega+1}{2}$, and
therefore for the EoS parameter in the range
$-1<\omega<-\frac{1}{3}$ the late time expansion will occur with a
positive acceleration.

In the case of a phantom fluid $\omega<-1$, the solution
corresponding to the expanding scale factor reads as \cite{Luc}
\begin{equation}\label{K1}
a_{ph}(t)\sim (t_{rip}-t)^{-\frac{2}{3|\omega+1|}}.
\end{equation}In this case $t$ is smaller than the constant
$t_{rip}$, and thus as $t\rightarrow t_{rip}$, the scale factor
diverges. The evolution of the Universe ends up with a finite-time
singularity, the so-called big-rip singularity \cite{Cal,Dab}.
Since the phantom energy density is proportional to the scale
factor as $\rho \sim a^{3|\omega+1|}$, at the big-rip the energy
density and pressure also diverge. One can easily see that this is
different from the ordinary big-crunch singularity at which the
energy density and pressure blow up as the scale factor tends to
zero at a finite time. In Refs. \cite{Dab} some exact solutions of
classical phantom cosmology were studied. These solutions show
that including a cosmological constant into the model gives rise
to a scale factor which its evolution, for standard types of
matter, begins with big-bang and terminates at big-crunch, while
for the phantom case, it begins with big-rip and terminates also
at a big-rip. An interesting remark of the phantom solution
(\ref{K1}) is that it looks to be dual of the standard matter
solution (\ref{J}) under the duality transformation
\begin{equation}\label{K2}
a\leftrightarrow \frac{1}{a},\hspace{0.5cm}\omega+1
\leftrightarrow |\omega+1|.\end{equation}Such duality is one of
the major features of the solutions of equations of motion in
phantom cosmology, so that if the set $(a, \omega+1)$ solves the
equations of motion for the standard matter models, the set
$(a^{-1}, |\omega+1|)$ solves the same equations for the phantom
model. In the following we shall see that how these classical
pictures will be modified if one takes into account the
quantum-mechanical considerations in the problem at hand.

\subsection{The quantum model}
We now focus attention on the study of the quantum cosmology of
the models described above. The momentum conjugate to $a$ is
\begin{equation}\label{L}
P_a=\frac{\partial {\cal L}}{\partial
\dot{a}}=-12N^{-1}a\dot{a},\end{equation}giving rise to the
following Hamiltonian
\begin{equation}\label{M}
H=N{\cal
H}=N\left[-\frac{P_a^2}{24a}+M_{\omega}a^{-3\omega}\right].\end{equation}We
now quantize the system with the use of the Wheeler-DeWitt
equation, that is, ${\cal H}\Psi=0$, where ${\cal H}$ is the
operator form of the Hamiltonian given by the above equation, and
$\Psi$ is the wave function of the Universe, a function of the
scale factor and the matter fields, if they exist. With the
replacement $P_a\rightarrow -i\frac{d}{da}$ we get the
Wheeler-DeWitt equation as
\begin{equation}\label{N}
\left(\frac{d^2}{da^2}+\frac{p}{a}\frac{d}{da}+24M_{\omega}a^{1-3\omega}\right)\Psi(a)=0,\end{equation}
where the parameter $p$ represents the ambiguity in the ordering
of factors $a$ and $P_a$ in the first term of (\ref{M}). For large
values of $a$, the solution of the above equation can be easily
obtained in the WKB (semiclassical) approximation. In this regime
we can neglect the second term in Eq. (\ref{N}). Then substituting
$\Psi(a)=\Omega(a)e^{iS(a)}$ in this equation leads to the
modified Hamilton-Jacobi equation
\begin{equation}\label{N1}
-\left(\frac{dS}{da}\right)^2+24M_{\omega}a^{1-3\omega}+{\cal
Q}=0,\end{equation}in which the quantum potential is defined as
${\cal Q}=\frac{1}{\Omega}\frac{d^2\Omega}{da^2}$. It is well
known that the quantum effects are important for small values of
the scale factor and in the limit of the large scale factor can be
neglected. Therefore, in the semiclassical approximation region we
can omit the ${\cal Q}$ term in Eq. (\ref{N1}) and obtain the
phase function $S(a)$ as
\begin{equation}\label{N2}
S=\pm
\frac{2\sqrt{24M_{\omega}}}{3(1-\omega)}a^{\frac{3(1-\omega)}{2}},\end{equation}where
the positive sign corresponds to an expanding Universe. In the WKB
method, the correlation between classical and quantum solutions is
given by the relation $P_a=\frac{\partial S}{\partial a}$. Thus,
using the definition of $P_a$ in (\ref{L}), the equation for the
classical trajectories becomes
\begin{equation}\label{N3}
-12a\dot{a}=\sqrt{24M_{\omega}}a^{\frac{1-3\omega}{2}},\end{equation}from
which one finds
\begin{equation}\label{N4}
a(t)\sim
\left(\sqrt{\frac{3M_{\omega}}{8}}(1+\omega)t\right)^{\frac{2}{3(1+\omega)}},\end{equation}where
shows that the late time behavior of the classical cosmology
(\ref{J}) is exactly recovered. The meaning of this result is that
for large values of the scale factor the effective action
corresponding to the expanding Universe is very large and the
Universe can be described classically. On the other hand, for
small values of the scale factor we cannot neglect the quantum
effects and the classical description breaks down. Since the WKB
approximation is no longer valid in this regime, one should go
beyond the semiclassical approximation. In general, the two
linearly independent solutions to Eq. (\ref{N}) can be expressed
in terms of the Bessel functions $J$ and $Y$ leading to the
following general solution
\begin{equation}\label{O}
\Psi(a)=a^{\frac{1-p}{2}}\left[c_1J_{\frac{1-p}{3(1-\omega)}}\left(\frac{4\sqrt{6M_{\omega}}}{3(1-\omega)}a^{\frac{3(1-\omega)}{2}}\right)+
c_2Y_{\frac{1-p}{3(1-\omega)}}\left(\frac{4\sqrt{6M_{\omega}}}{3(1-\omega)}a^{\frac{3(1-\omega)}{2}}\right)\right].\end{equation}Note
that the minisuperspace of the above model has only one degree of
freedom denoted by the scale factor $a$ in the range
$0<a<+\infty$. According to \cite{Vil}, its nonsingular boundary
is the line $a=0$, while at the singular boundary this variable is
infinite. Now, we impose the boundary condition on the above
solutions such that at the nonsingular boundary the wave function
vanishes \cite{Vil}, which yields $c_2=0$, and thus we arrive at
the unique solution
\begin{equation}\label{P}
\Psi(a)=a^{\frac{1-p}{2}}J_{\frac{1-p}{3(1-\omega)}}\left(\frac{4\sqrt{6M_{\omega}}}{3(1-\omega)}a^{\frac{3(1-\omega)}{2}}\right).\end{equation}Note
that Eq. (\ref{N}) is a Schr\"{o}dinger-like equation for a
fictitious particle with zero energy moving in the field of the
superpotential $U(a)=-24M_{\omega}a^{1-3\omega}$. Usually, in the
presence of such a potential the minisuperspace can be divided
into two regions, $U>0$ and $U<0$, which could be termed the
classically forbidden and classically allowed regions
respectively. In the classically forbidden region the behavior of
the wave function is exponential while in the classically allowed
region the wave function behaves oscillatory. In the quantum
tunneling approach \cite{Vil}, the wave function is so constructed
as to create a Universe emerging from {\it nothing} by a tunneling
procedure through a potential barrier in the sense of usual
quantum mechanics. Now, in our model the superpotential is always
negative which means that there is no possibility of tunneling
anymore since a zero energy system is always above the
superpotential. In such a case tunneling is no longer required as
classical evolution is possible. As a consequence the wave
function always exhibits oscillatory behavior.
\begin{figure}
\begin{tabular}{c} \epsfig{figure=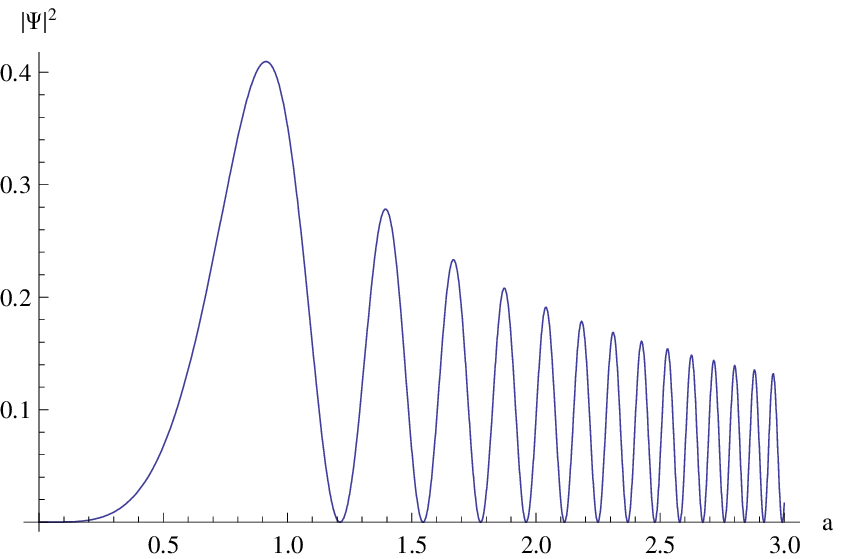,width=4.5cm}
\hspace{1cm} \epsfig{figure=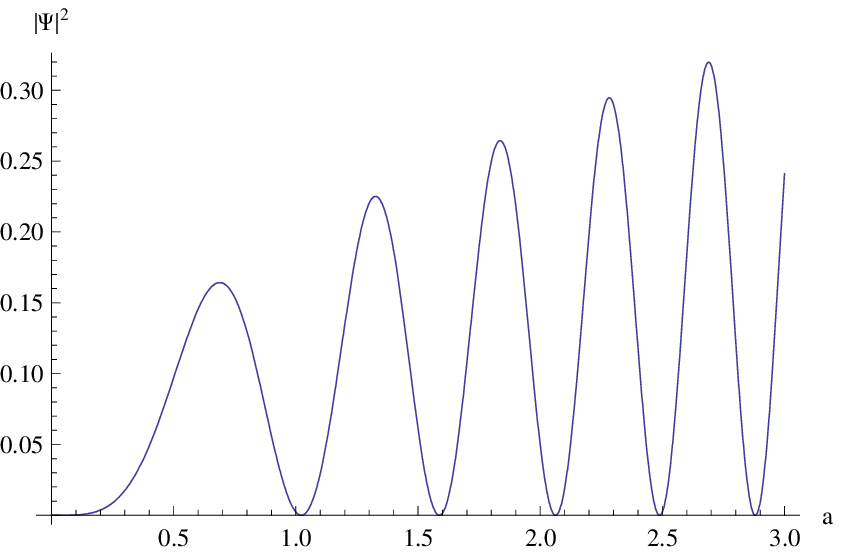,width=4.5cm}\hspace{1cm}
\epsfig{figure=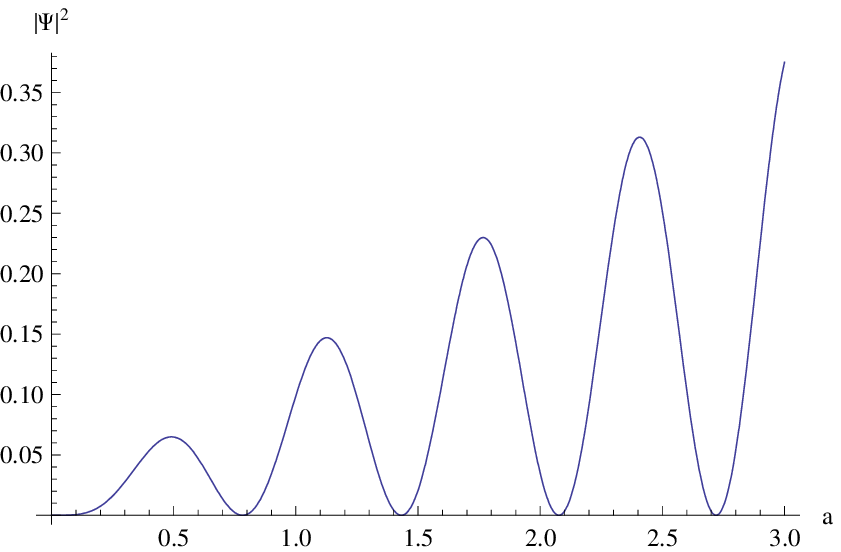,width=4.5cm}
\end{tabular}
\caption{\footnotesize The square of the wave function for the
quantum Universe. We take the numerical values $M_{\omega}=1$,
$p=-1$ and $\omega=-1,0,1/3$.}\label{fig1}
\end{figure}
In figure \ref{fig1} we have plotted the square of the wave
functions for typical values of the parameters. It is seen from
this figure that the wave function has a well-defined behavior
near $a=0$ and describes a Universe emerging out of {\it nothing}
without any tunneling. (See \cite{Lin} in which such a wave
functions are also appeared in the case study of the probability
of quantum creation of compact flat and open de Sitter (dS)
Universes.) On the other hand, the emergence of several peaks in
the wave function may be interpreted as a representation of
different quantum states that may communicate with each other
through tunneling. This means that there are different possible
Universes (states) from which our present Universe could have
evolved and tunneled in the past, from one Universe (state) to
another.
\section{Gauge-fixed perfect fluid cosmology}
In this section we deal with the problem at hand in an another
point of view. As is well known from the classical mechanics, the
Lagrangian of a given dynamical system is unique up to a total
derivative of time. This means that if $L$ is a Lagrangian for a
dynamical system satisfying Lagrange's equations then
$L'=L+\frac{df}{dt}$ also satisfies Lagrange's equations, where
$f$ is a differentiable function of coordinates and time.
Therefore, with different choices of the function $f$ we obtain
different equivalent Lagrangians which are called trivial or gauge
equivalent Lagrangians. On the other hand, for such a dynamical
system one can demonstrate that very different Lagrangians lead to
the same equation of motion. For instance, for a one dimensional
particle moving under the act of potential $V(x)$, in addition of
the usual Lagrangian $L=\frac{1}{2}m\dot{x}^2-V(x)$ and its gauge
equivalent, one can show that the Lagrangian
$L=\frac{1}{12}m^2\dot{x}^4+m\dot{x}^2V(x)-V^2(x)$ also gives the
correct equation of motion \cite{Gold}. The same is true for the
Lagrangians $L=\frac{1}{n}\dot{x}^n$ ($n\geq 2$) for a one
dimensional free particle and $L=T-V+\frac{\gamma J}{r^2}$
($\gamma$ is a constant and $J$ is the angular momentum) for a
particle moving under a spherically symmetric potential. The
problem of finding these so-called nontrivial equivalent
Lagrangians is the subject of the inverse problem in the calculus
of variation \cite{Mor}. Although from the (nontrivial) equivalent
Lagrangians we obtain the same classical equations of motion,
canonical quantization of the system under consideration based on
these Lagrangians may yield different results. Hence, classical
equivalent systems may be nonequivalent quantum mechanically. In
nonlinear systems (like gravitational field equations) one can
still think about another kinds of Lagrangians which give rise to
the correct dynamics for the system while the equations of motion
may be different with the ones coming from the usual Lagrangian.
In the next subsection we shall introduce such a Lagrangian for
our model at hand and see that although its equation of motion is
somehow different with once of Einstein-Hilbert action, the
solutions are the same. In this sense we refer these kinds of
Lagrangians also as the non trivial equivalent with the
traditional Einstein-Hilbert Lagrangian.

\subsection{The classical model}
Now, we consider again the flat FRW Universe (\ref{A}) filled with
a perfect fluid with EoS (\ref{C}). Instead of the consensus
Lagrangian (\ref{D}) we examine the following (up to a overall
factor) Lagrangian proposed in \cite{Mal}
\begin{equation}\label{Q}
{\cal L}=a^{3\omega+1}\dot{a}^2.\end{equation}A simple calculation
based on the Euler-Lagrange equation gives the equation of motion
as
\begin{equation}\label{R}
(3\omega+1)\dot{a}^2+2a\ddot{a}=0,\end{equation}where its solution
reads
\begin{equation}\label{S}
a(t)={\cal A}\left[3(\omega+1)t-{\cal
B}\right]^{\frac{2}{3(\omega+1)}},\end{equation}for $\omega \neq
-1$, and
\begin{equation}\label{S1}
a(t)={\cal C}_1e^{{\cal C}_2t},\end{equation}for $\omega =-1$. We
may set the integration constants ${\cal A}$, ${\cal B}$, ${\cal
C}_1$ and ${\cal C}_2$ as ${\cal
A}=\left(\frac{M_{\omega}}{24}\right)^{\frac{1}{3(\omega+1)}}$,
${\cal B}=\tau_0 \left(\frac{24}{M_{\omega}}\right)^{1/2}$, ${\cal
C}_1=a_0$ and ${\cal C}_2=\sqrt{\frac{M_{-1}}{6}}$ so that the
above solutions coincide with the expressions (\ref{J}) and
(\ref{J1}) for the corresponding classical cosmology. Therefore,
although the difference between the Lagrangians (\ref{Q}) and
(\ref{D}) is not a total derivative of time and these two
Lagrangians yield different classical equations of motion
(\ref{E}) and (\ref{R}), their equations of motion have the same
solutions. As mentioned above, we refer the Lagrangian (\ref{Q})
as an (nontrivial) equivalent Lagrangian for the system under
consideration. Now, the question is that what the relation between
two Lagrangians is? As is well known, general relativity is a
parameterized gauge theory which its canonical formalism is based
on the Hamiltonian and momentum constraints. Because of the
existence of these constraints the quantum version of this theory
suffers from a number of problems, namely the construction of the
Hilbert space to define a positive definite inner product of the
solutions of the  Wheeler-DeWitt equation, the operator ordering
problem, and also, most importantly, the problem of time. As we
have seen in the previous section, the wave function in the
Wheeler-DeWitt equation is independent of time, i.e., the Universe
has a static picture in this scenario. This problem was first
addressed in \cite{Dew} by DeWitt himself. However, he argued that
the problem of time should not be considered as a hindrance in the
sense that the theory itself must include a suitable well-defined
time in terms of its geometry or matter fields. In this scheme
time is identified with one of the characters of the geometry or
with a scalar character of matter fields coupled to gravity in any
specific model. Identification of time with one of the dynamical
variables depends on the method we use to deal with the
constraints. In any constrained system we can impose the
constraints in different steps. In classical mechanics, for
example, we may first solve the equations of constraint to reduce
the degrees of freedom of the system and obtain a minimal number
of dynamical variables. On the other hand, we may multiply the
constraint by a variable parameter and add it to the Lagrangian.
This Lagrange multiplier plays the role of an additional dynamical
variable and the equations of motion consist of those obtained
from variation of the Lagrangian with respect to the dynamical
variables plus the equation of constraint. Also, when quantizing
the system, we may impose the constraint before or after the
quantization has been done. Now, if our system is the entire
Universe, e.g., in the case of quantum cosmology, these procedures
result in different approaches to the problem of time
reparameterization \cite{Is}.

In writing Lagrangian (\ref{Q}), as is argued in \cite{Mal} the
gauge freedom is removed at the level of Lagrangian and the system
is reduced to its true degree of freedom. Hence, the Hamiltonian
of the model is not expected to vanish identically. Therefore, in
quantization of such a system we should deal with a
Schr\"{o}dinger equation instead of the  Wheeler-DeWitt equation
in which we have a time dependent wave function.

Let us now set up the Hamiltonian formalism of the theory based on
the Lagrangian (\ref{Q}). We introduce the conjugate momentum as
\begin{equation}\label{T}
P_a=\frac{\partial {\cal L}}{\partial
\dot{a}}=2a^{3\omega+1}\dot{a},\end{equation}and by the usual
Legendre transformation we obtain the Hamiltonian as
\begin{equation}\label{U}
{\cal H}=\frac{P_a^2}{4a^{3\omega+1}}.\end{equation}It is easy to
see that the classical Hamiltonian equations admit the solution
(\ref{S}). Now, if using (\ref{S}), we compute the numerical value
of the Hamiltonian (\ref{U}) we get
\begin{equation}\label{V}
{\cal H}=4{\cal A}^{3(\omega+1)}=E=\mbox{const}.\end{equation}As
we have mentioned above, in contrast to the Hamiltonian constraint
in parameterized theory in the previous section, in the
gauge-fixed theory based on the equivalent Lagrangian the
corresponding Hamiltonian is not identically equal to zero.

\subsection{The quantum model}
To pass to the quantum version of this model we should note that
the usual approach to canonical quantization of a cosmological
model is the Wheeler-DeWitt approach where one uses the Dirac
method to quantize the degrees of freedom of the system. The role
of constraints in their operator form is to annihilate the wave
function of the Universe. This procedure leads one to the basic
equation of quantum cosmology, the so-called  Wheeler-DeWitt
equation. This is what we have performed in the previous section.
However, as was done in this section, one may fix the constraint
before using it in the theory, in particular before quantizing the
system. If we do so, we are led to the Schr\"{o}dinger equation
\begin{equation}\label{W}
{\cal H}\Psi(a,t)=i\frac{\partial}{\partial
t}\Psi(a,t),\end{equation}where ${\cal H}$ is the operator form of
the reduced Hamiltonian (\ref{U}) and $\Psi(a,t)$ is the time
dependent wave function of the Universe. With the usual
replacement $P_a\rightarrow -i\partial_a$ and choice of a
particular factor ordering this equation becomes
\begin{equation}\label{X}
\frac{1}{4}\left(\frac{3\omega+1}{2}a^{-3\omega-2}\frac{\partial}{\partial
a}-a^{-3\omega-1}\frac{\partial^2}{\partial
a^2}\right)\Psi(a,t)=i\frac{\partial}{\partial
t}\Psi(a,t).\end{equation}We separate the variables in the
Schr\"{o}dinger Eq. (\ref{X}) as
\begin{equation}\label{Y}
\Psi(a,t)=e^{-iEt}\psi(a),\end{equation}leading to
\begin{equation}\label{Z}
\left(\frac{3\omega+1}{2}a^{-3\omega-2}\frac{d}{da}-a^{-3\omega-1}\frac{d^2}{da^2}-4E\right)\psi(a)=0.\end{equation}For
$\omega \neq -1$ the solutions of the above differential equation
may be written in the form
\begin{equation}\label{AB}
\psi(a)=d_1\sin\left[\frac{4\sqrt{E}}{3(\omega+1)}a^{\frac{3(\omega+1)}{2}}\right]+d_2\cos\left[\frac{4\sqrt{E}}{3(\omega+1)}a^{\frac{3(\omega+1)}{2}}\right],
\end{equation}where $d_1$ and $d_2$ are integration constants. If
we impose the boundary condition $\psi(a=0)=0$ on these solutions,
the eigenfunctions of the Schr\"{o}dinger equation can be written
as
\begin{equation}\label{AC}
\Psi_E(a,t)=e^{-iEt}\sin\left[\frac{4\sqrt{E}}{3(\omega+1)}a^{\frac{3(\omega+1)}{2}}\right].\end{equation}We
may now write the general solution to the Schr\"{o}dinger equation
as a superposition of its eigenfunctions, that is
\begin{equation}\label{AD}
\Psi(a,t)=\int_0^\infty A(E)\Psi_E(a,t)dE,\end{equation}where
$A(E)$ is a suitable weight function to construct the wave
packets. By using the equality \cite{Ab}
\begin{equation}\label{AE}
\int_0^\infty e^{-\gamma x}\sin \sqrt{mx}dx=\frac{\sqrt{\pi
m}}{2\gamma^{3/2}}e^{-\frac{m}{4\gamma}},\end{equation}we can
evaluate the integral over $E$ in (\ref{AD}) and simple analytical
expression for this integral is found if we choose the function
$A(E)$ to be a quasi-Gaussian weight factor $A(E)=e^{-\gamma E}$
($\gamma$ is an arbitrary positive constant), which results in
\begin{equation}\label{AF}
\Psi(a,t)=\int_0^\infty e^{-\gamma
E}e^{-iEt}\sin\left[\frac{4\sqrt{E}}{3(\omega+1)}a^{\frac{3(\omega+1)}{2}}\right]dE.\end{equation}Using
of the relation (\ref{AE}) leads to the following expression for
the wave function
\begin{equation}\label{AG}
\Psi(a,t)={\cal
N}\frac{a^{\frac{3(\omega+1)}{2}}}{(\omega+1)(\gamma+it)^{3/2}}\exp
\left[-\frac{4a^{3(\omega+1)}}{9(\omega+1)^2(\gamma+it)}\right],\end{equation}where
${\cal N}$ is a numerical factor. Now, having the above expression
for the wave function of the Universe, we are going to obtain the
predictions for the behavior of the dynamical variables in the
corresponding cosmological model. To do this, in the next
subsection, we shall adopt two approaches to evaluate the
classical behavior of the dynamical variables in the model which
lead to the same results. In the many-worlds interpretation of
quantum mechanics \cite{Tip}, we can calculate the expectation
values of the dynamical variables and, in the realm of the
ontological interpretation of quantum mechanics \cite{Boh}, one
can evaluate the Bohmian trajectories for those variables.
\begin{figure}
\begin{tabular}{c} \epsfig{figure=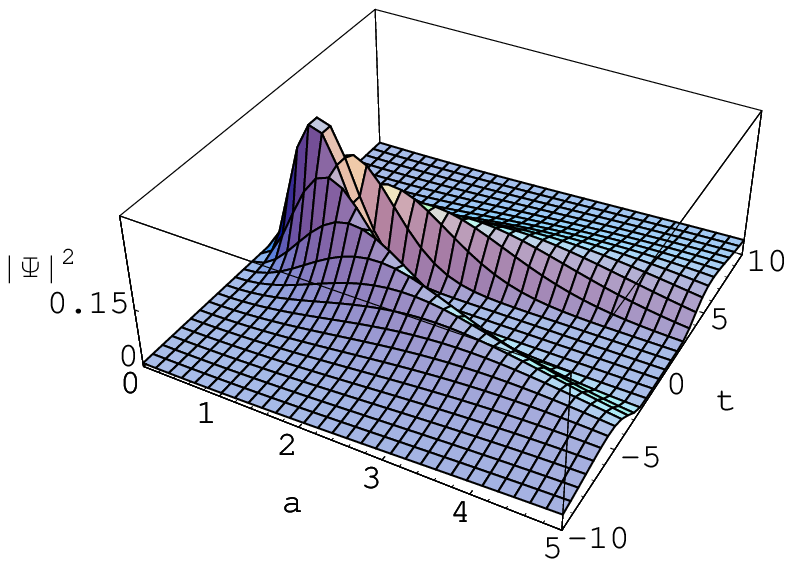,width=4.5cm}
\hspace{1cm} \epsfig{figure=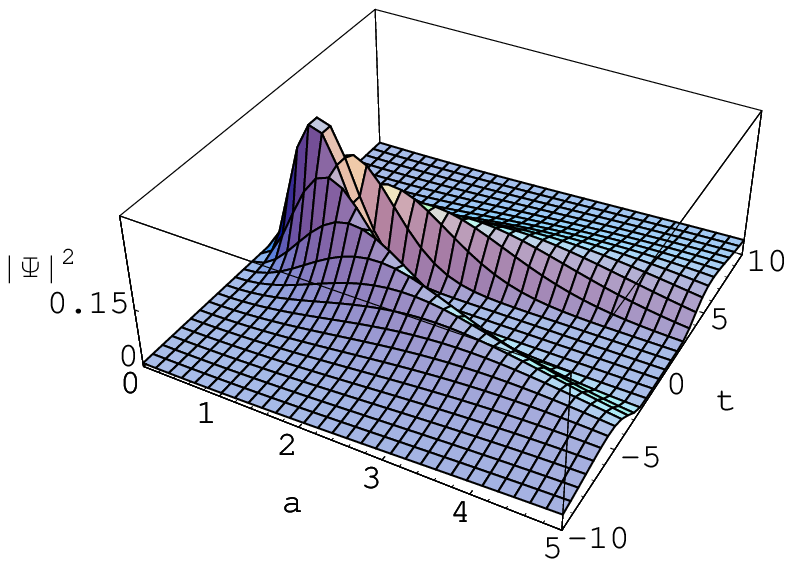,width=4.5cm}\hspace{1cm}
\epsfig{figure=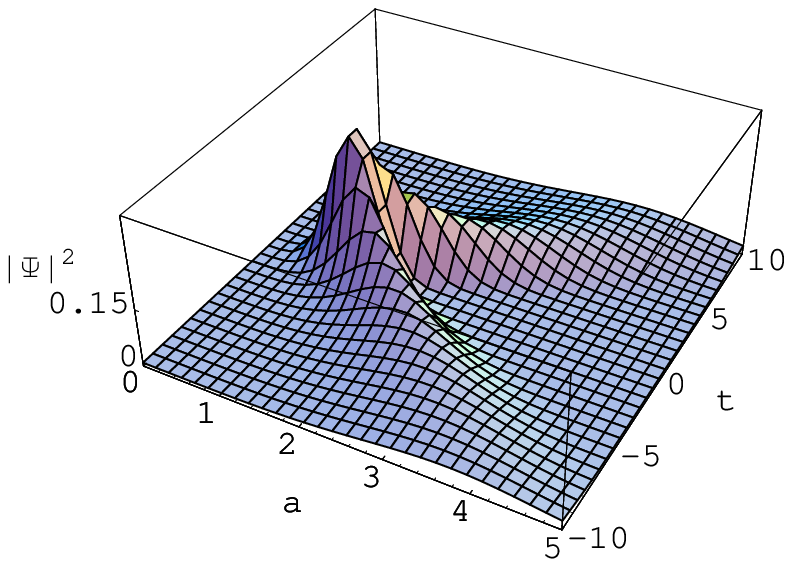,width=4.5cm}
\end{tabular}
\caption{\footnotesize The figures show $|\Psi(a,t)|^2$, the
square of the wave function. The figures are plotted for the
numerical values ${\cal N}=\gamma=1$, and we set the equation of
state parameter (from left to right) $\omega=-1/3,0,1/3$. After
examining some other values for this parameter, we verify that the
general behavior of the wave function is repeated.}\label{fig2}
\end{figure}
In figure \ref{fig2} we have plotted the square of the wave
function for typical numerical values of the parameters. As this
figure shows, at $t=0$, the wave function has a dominant peak in
the vicinity of some nonzero value of $a$. This means that the
wave function predicts the emergence of the Universe from a
nonsingular state corresponding to this dominant peak. As time
progresses, the wave packet begins to propagate in the $a$
direction, its width becoming wider and its peak moving towards
the greater values of $a$. The wave packet disperses as time
passes, the minimum width being attained at $t=0$. As in the case
of the free particle in quantum mechanics, the more localized the
initial state at $t=0$, the more rapidly the wave packet
disperses. Therefore, the quantum effects make themselves felt
only for small enough $t$ corresponding to small $a$ as expected,
and the wave function predicts that the Universe will assume
states with larger $a$ in its late time evolution.

\begin{figure}
\begin{tabular}{c} \epsfig{figure=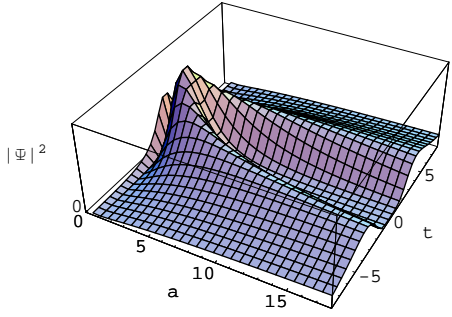,width=6cm}
\end{tabular}
\caption{\footnotesize The figure shows $|\Psi(a,t)|^2$, the
square of the wave function for $\omega=-1$. The figure is plotted
for the numerical values ${\cal N}=1$ and
$\gamma=1.5$.}\label{fig3}
\end{figure}

The above solutions are not valid for $\omega=-1$. In this
particular case, Eq. (\ref{Z}) becomes an Euler's type equation,
and following the steps (\ref{AB})-(\ref{AF}) the final result for
the wave function takes the form
\begin{equation}\label{AG1}
\Psi(a,t)={\cal N}\frac{\ln a}{(\gamma+it)^{3/2}}\exp
\left[-\frac{\ln^2a}{\gamma+it}\right].\end{equation}The behavior
of this function is represented in figure \ref{fig3}. The
discussions on the comparison between quantum cosmological
solution and its classical counterpart are the same as previous
models, namely the $\omega \neq -1$ models. Similar discussion as
above would be applicable to this case as well.

\subsection{Recovery of the classical solutions}
In general, one of the most important features in quantum
cosmology is the recovery of classical cosmology from the
corresponding quantum model or, in other words, how can the
quantum wave functions predict a classical Universe. In this
approach, one usually constructs a coherent wave packet with good
asymptotic behavior in the minisuperspace, peaking in the vicinity
of the classical trajectory. On the other hand, in an another
approach to show the correlations between classical and quantum
pattern, following the many-worlds interpretation of quantum
mechanics \cite{Tip}, one may calculate the time dependence of the
expectation value of a dynamical variable $q$ as
\begin{equation}\label{AH}
<q>(t)=\frac{<\Psi|q|\Psi>}{<\Psi|\Psi>}.\end{equation}Following
this approach, we may write the expectation value for the scale
factor as
\begin{equation}\label{AI}
<a>(t)=\frac{\int_0^\infty \Psi^*(a,t)a\Psi(a,t)da}{\int_0^\infty
\Psi^*(a,t)\Psi(a,t)da},\end{equation}which yields
\begin{equation}\label{AJ}
<a>(t)=\left(\frac{9}{8}\frac{(\omega+1)^2}{\gamma}\right)^{\frac{1}{3(\omega+1)}}\frac{\Gamma\left(\frac{5+3\omega}{3(1+\omega)}\right)}
{\Gamma\left(\frac{4+3\omega}{3(1+\omega)}\right)}\left(\gamma^2+t^2\right)^{\frac{1}{3(1+\omega)}}.\end{equation}It
is important to classify the nature of the quantum model as
concerns the presence or absence of singularities. For the wave
function (\ref{AG}), the expectation value (\ref{AJ}) of $a$ never
vanishes, showing that these states are nonsingular. Indeed, in
(\ref{AJ}) $t$ varies from $-\infty$ to $+\infty$ and $t=0$ is
just a specific moment without any particular physical meaning
like big-bang singularity. The expression (\ref{AJ}) for $\omega
\neq -1$, represents a bouncing Universe with no singularity where
its late time behavior coincides to the late time behavior of the
classical solution (\ref{J}), that is $a(t)\sim
t^{\frac{2}{3(\omega+1)}}$ . Now we can calculate the dispersion
of the wave packet in the $a$ direction, which is defined as
\begin{equation}\label{AL}
(\Delta a)^2=<a^2>-<a>^2,\end{equation}using (\ref{AG}) and
(\ref{AJ}), we get
\begin{equation}\label{AM}
(\Delta a)^2 \sim
\left(\gamma^2+t^2\right)^{\frac{2}{3(1+\omega)}}.\end{equation}The
result is that the wave packet traveling in the $a$ direction
spreads as time increases and thus its degree of localization is
reduced. The width of the wave packet evaluated in (\ref{AM})
agrees with the discussion in the end of the previous subsection.
Indeed, we may interpret the above relation for the width of the
wave function as the coincidence of the classical trajectories
with the quantum ones for large values of time. Therefore, in view
of the behavior of the scale factor, the classical solution
(\ref{AJ}) is in complete agreement with the quantum patterns
shown in figure \ref{fig2}, and both predict a (nonsingular)
monotonically increasing evolution for the scale factor and
consequently there is an almost good correlation between the
quantum patterns and classical trajectories.

The issue of the correlation between classical and quantum schemes
may be addressed from another point of view. It is known that the
results obtained by using the many-world interpretation agree with
those that can be obtained using the ontological interpretation of
quantum mechanics \cite{Boh}. In Bohmian interpretation, the wave
function is written as
\begin{equation}\label{AN}
\Psi(a,t)=\Omega(a,t)e^{iS(a,t)},\end{equation}where $\Omega$ and
$S$ are some real functions. Substitution of this expression into
the Schr\"{o}dinger Eq. (\ref{X}) leads to the continuity equation
\begin{equation}\label{AP}
\frac{1}{4}\left[\frac{3\omega+1}{2}a^{-3\omega-2}\Omega\frac{\partial
S}{\partial a}-2a^{-3\omega-1}\frac{\partial S}{\partial
a}\frac{\partial \Omega}{\partial
a}-a^{-3\omega-1}\frac{\partial^2 S}{\partial
a^2}\right]-\frac{\partial \Omega}{\partial t}=0,\end{equation}and
the modified Hamilton-Jacobi equation
\begin{equation}\label{AO}
\frac{\partial S}{\partial
t}+\frac{1}{4}a^{-3\omega-1}\left(\frac{\partial S}{\partial
a}\right)^2+{\cal Q}=0,\end{equation}in which the quantum
potential ${\cal Q}$ is defined as
\begin{equation}\label{AR}
{\cal
Q}=\frac{1}{4\Omega}\left[\frac{3\omega+1}{2}a^{-3\omega-2}\frac{\partial
\Omega}{\partial a}-a^{-3\omega-1}\frac{\partial^2\Omega}{\partial
a^2}\right].\end{equation}From the wave function (\ref{AG}), the
real functions $\Omega(a,t)$ and $S(a,t)$ can be obtained as
\begin{equation}\label{AS}
\Omega(a,t)={\cal
N}\frac{a^{\frac{3(\omega+1)}{2}}}{(\omega+1)(\gamma^2+t^2)^{3/4}}\exp\left[-\frac{4\gamma}{9(\omega+1)^2(\gamma^2+t^2)}a^{3(\omega+1)}\right],
\end{equation}
and
\begin{equation}\label{AT}
S(a,t)=-\frac{3}{2}\arctan
\frac{t}{\gamma}+\frac{4t}{9(\omega+1)^2(\gamma^2+t^2)}a^{3(\omega+1)}.\end{equation}In
this interpretation the classical trajectories, which determine
the behavior of the scale factor are given by $P_a=\frac{\partial
S}{\partial a}$. Using the expression for $P_a$ in (\ref{T}), the
equation for the classical trajectories becomes
\begin{equation}\label{AU}
2a^{3\omega+1}\dot{a}=\frac{4t}{3(\omega+1)(\gamma^2+t^2)}a^{3\omega+2}.\end{equation}Therefore,
after integration we get
\begin{equation}\label{AV}
a(t)=a_0\left(\gamma^2+t^2\right)^{\frac{1}{3(\omega+1)}},\end{equation}where
$a_0$ is a constant of integration. This solution has the same
behavior as the expectation value computed in (\ref{AJ}) and like
that is free of singularity. The origin of the singularity
avoidance may be understood by the existence of the quantum
potential which corrects the classical equations of motion.
Inserting the relation (\ref{AV}) in (\ref{AS}) and then using
(\ref{AR}), we can find the quantum potential in terms of the
scale factor as
\begin{equation}\label{AW}
{\cal Q}\sim a^{-3(\omega+1)}.\end{equation}It is obvious from
this equation that the quantum effects are important for small
values of the scale factor and in the limit of the large scale
factor can be neglected. Therefore, asymptotically the classical
behavior is recovered. In this sense we can extract a repulsive
force from the quantum potential (\ref{AW}) as
\begin{equation}\label{BA}
F_a=-\frac{\partial {\cal Q}}{\partial a}\sim
a^{-(3\omega+4)},\end{equation}which may interpreted as being
responsible of the avoidance of singularity. For small values of
$a$ (near the big-bang singularity), this repulsive force takes a
large magnitude and thus prevents the scale factor to evolve to
the classical singularity $a\rightarrow 0$.

Again, for the case where the perfect fluid plays the role of a
cosmological constant, i.e. $\omega=-1$, using the wave function
(\ref{AG1}) the above steps lead us the following Bohmian  value
for the scale factor
\begin{equation}\label{BC}
a(t)\sim e^{\sqrt{\gamma^2+t^2}}.\end{equation}In comparison with
the classical solutions, it is seen that this model represents  a
bouncing cosmology which recovers the late time behavior of the
classical de Sitter Universe (\ref{S1}). In this case the quantum
potential takes the form ${\cal Q}\sim \frac{1}{\ln^2a}$, which
yields a repulsive force $F_a\sim \frac{1}{a\ln^3a}$. For
$a\rightarrow 0$ (in de Sitter model this occurs in the limit
$t\rightarrow -\infty$) $F_a$ takes a large magnitude and thus
like the above discussion this force may interpreted as being
responsible of the avoidance of this kind of singularity.

To verify how this formalism work for the case of the phantom
fluid, we use the duality transformation (\ref{K2}). In this case
it is easy to show that the phantom counterpart of the solution
(\ref{AV}) reads
\begin{equation}\label{BD}
a_{ph}(t)\sim
\left[\gamma^2+(t_{rip}-t)^2\right]^{-\frac{1}{3|\omega+1|}},\end{equation}in
which for comparison with the classical solution, we absorb the
integration constants in $t_{rip}$. We see that although this
solution has the same limiting behavior as the classical phantom
scale factor (\ref{K1}), its behavior near $t_{rip}$ is different.
Indeed, while in the classical solution (\ref{K1}), which is valid
only in the range $t<t_{rip}$, the scale factor increase
monotonically towards the big-rip singularity, its quantum version
(\ref{BD}) has a regular behavior at $t=t_{rip}$. This means that
the scale factor (\ref{BD}) is free of big-rip singularity at
$t=t_{rip}$ and at this moment the expansion behavior of the
Universe will be replaced by a contraction phase. The avoidance of
the big-rip singularity may be interpreted again by the existence
of the quantum potential. Using the duality (\ref{K2}) the quantum
potential for the phantom model can be written as
\begin{equation}\label{BE}
{\cal Q}_{ph}\sim a^{3|\omega+1|}.\end{equation}Now, we see that
the quantum effects show their important role for the large values
of the scale factor. Therefore, the repulsive force resulting from
this quantum potential (the phantom counterpart of (\ref{BA}))
will take its large magnitude for large values of $a$ and thus the
evolution of the scale factor towards the big-rip singularity will
be avoided\footnote{Here, we have taken only a quick look at the
quantum phantom model from a phenomenological point of view. In a
more fundamental study one may replace the phantom by a scalar
field with negative kinetic energy. In this sense the quantum
phantom cosmology may be investigated in the framework of scalar
field models, see\cite{DaKi}.}.
\begin{figure}
\begin{tabular}{c} \epsfig{figure=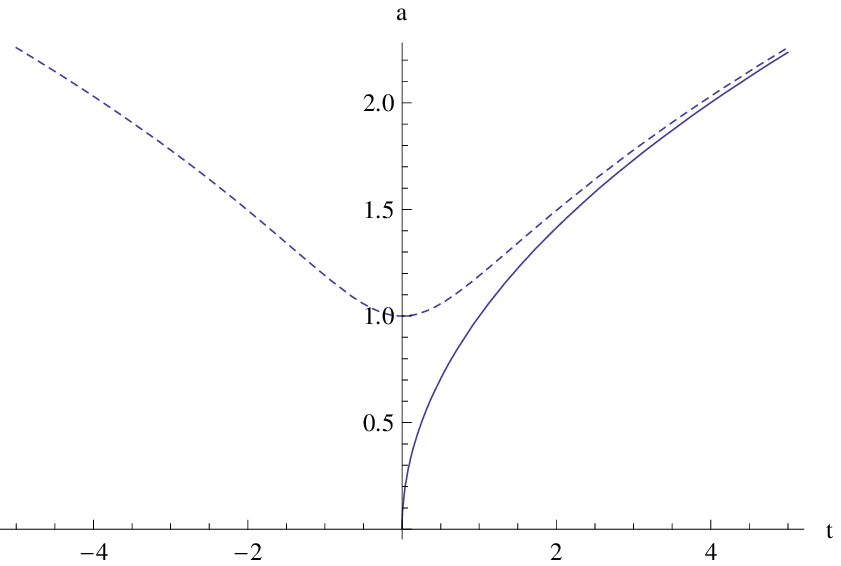,width=4.5cm}
\hspace{1cm} \epsfig{figure=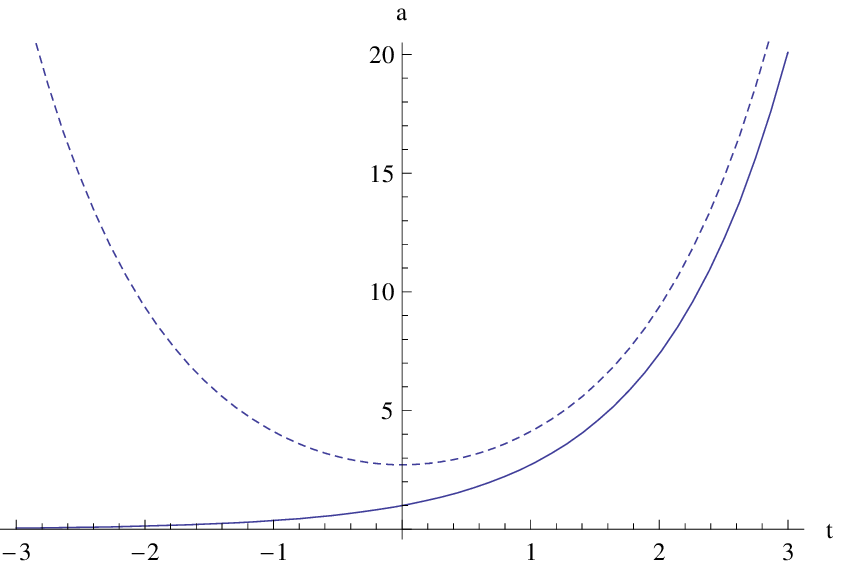,width=4.5cm}\hspace{1cm}
\epsfig{figure=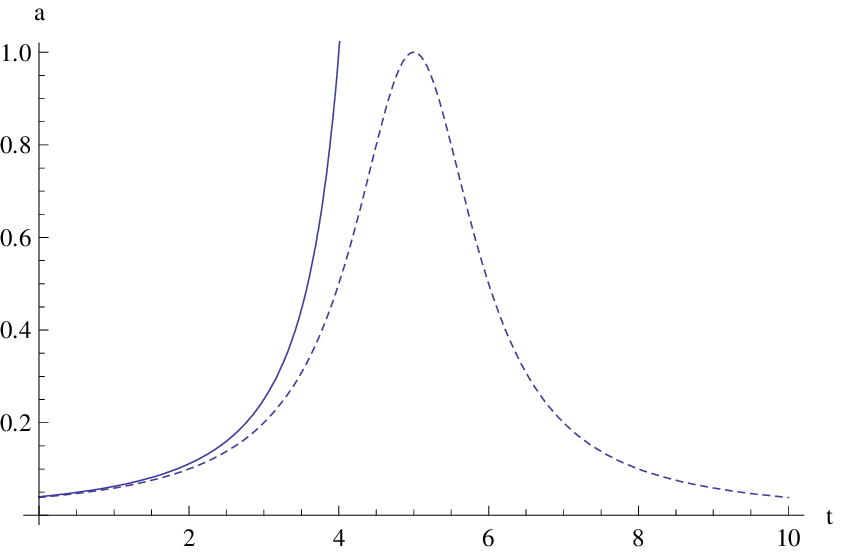,width=4.5cm}
\end{tabular}
\caption{\footnotesize Qualitative behavior of the classical scale
factor (solid lines) and its Bohmian or quantum expectation value
counterpart (dashed lines). The figures are plotted (from left to
right) for $\omega=1/3$, $\omega=-1$ and $\omega=-4/3$. Note that
for the models that exhibit the classical big-bang singularity,
(e.g. the left figure), while there is no classical behavior for
the negative values of $t$, in the quantum counterpart of the
scale factor $t$ can vary from $-\infty$ to $+\infty$, see
(\ref{AJ}) and (\ref{AV}).}\label{fig4}
\end{figure}

In figure \ref{fig4} we have summarized the results of this
subsection. As is clear from this figure, for the standard
cosmological fluids which we have chosen the cases $\omega=1/3$
(radiation) and $\omega=-1$ (cosmological constant) here, in
conventional inflationary cosmology, the Universe begins with a
singularity and expands forever. On the other hand, the evolution
of the scale factor based on the quantum-mechanical considerations
shows a bouncing behavior in which the Universe bounces from a
contraction epoch to a reexpansion era. We see that in the late
time of cosmic evolution in which the quantum effects are
negligible these two behaviors coincide with each other. This
means that the quantum structure which we have constructed based
on the gauge-fixed Lagrangian (\ref{Q}) has a good correlation
with its classical counterpart. Also, for the phantom case
$\omega=-4/3$, the singular behavior of the classical scale factor
at $t=t_{rip}$ is replaced by a transition from an expansion epoch
to a recontraction regime. With a glance at this figure one can
find the duality between the standard matter solutions (the figure
on the left) and the phantom solutions (the figure on the right).
This duality may be expressed in terms of the following
statements: \newline $\bullet$ Standard types of matter: There is
no classical behavior before the big-bang singularity. The quantum
effects dominate in the region of the classical big-bang
singularity, i.e., at the small values of scale factor. At the
big-bang the quantum solutions bounce from a contraction phase to
an expansion era.\newline $\bullet$ Phantom solutions: There is no
classical behavior after the big-rip singularity. The quantum
effects dominate in the region of the classical big-rip
singularity, i.e., at the large values of scale factor. At the
big-rip the quantum solutions fall from an expansion phase to a
contraction era.

As a final remark, we would like to emphasize that in comparison
with the quantum cosmological model presented in the previous
section, one of the advantages of using this formalism in our
quantum cosmological model is that, in a natural way, it can offer
a time parameter in terms of which we can evaluate dynamical
behavior of the cosmic scale factor.
\section{Conclusions}
In this paper we have studied the classical and quantum dynamics
of a perfect fluid cosmological model with an eye to the problem
of time gauge in quantum cosmology. To do this, in addition of the
Einstein-Hilbert action, we have examined a gauge-fixed Lagrangian
which turns out to correspond to a specific choice of time
parameter. We have seen that the classical evolution of the
Universe based on these two methods has the same solutions
represent a late time power law expansion coming from a big-bang
singularity when $-1<\omega<1$ (except in the case $\omega=-1$ for
which we had a de Sitter Universe) and tending to a big-rip
singularity when $\omega <-1$. In this sense the selected time
parameter in the gauge-fixed method is nothing but the traditional
cosmic time in the Einstein-Hilbert Lagrangian. We then dealt with
the quantization of the model in which we saw that the classical
singular behavior will be modified. In the consensus quantum
model, we showed that the Wheeler-DeWitt wave function has a good
semiclassical behavior and for large scale factor recovers the
classical solutions. Also, for small values of scale factor this
wave function shows a pattern in which there are several possible
quantum states from which our present Universe could have evolved
and tunneled in the past from one state to another. On the other
hand, using the gauge-fixed representation at the Lagrangian
level, we showed that it may lead to a Schr\"{o}dinger equation
for the quantum-mechanical description of the model. We showed
that the Schr\"{o}dinger equation can be separated and its
eigenfunctions can be obtained in terms of analytical functions.
By an appropriate superposition of the eigenfunctions, we
constructed the corresponding wave packet. The time evolution of
this wave packet represents its motion along the larger $a$
direction. As time passes, our results indicated that the wave
packets disperse and the minimum width being attained at $t=0$,
which means that the quantum effects are important for small
enough $t$ , corresponding to small $a$. The avoidance of
classical singularities due to quantum effects, and the recovery
of the classical dynamics of the Universe are another important
features of our quantum presentation of the model. These questions
have been investigated by two different methods. The time
evolution of the expectation value of the dynamical variables and
also their Bohmian counterparts have been evaluated in the spirit
of the many-worlds and ontological interpretation of quantum
cosmology respectively. We verified that a bouncing
singularity-free Universe is obtained in both cases. The use of
the ontological interpretation has allowed us to understand the
origin of the avoidance of singularity by a repulsive force due to
the existence of the quantum potential. The repulsive nature of
this force prevents the Universe to reach the singularity. We have
also taken a glance at the quantum phantom model in this framework
and showed that, in this case, the quantum effects dominate in the
region of the classical big-rip singularity, i.e., the quantum
effects occur at large values of the scale factor. We saw that the
quantum-mechanical considerations in the phantom model result in a
scale factor which has a regular behavior at the big-rip moment,
changing its evolution from an expansion era to a contraction
epoch. Therefore, the big-rip singularity may be removed in the
quantum theory.

\end{document}